\documentclass[11pt,a4paper]{amsart}
\usepackage{amsmath,amsthm,amscd,amssymb,latexsym,graphicx,upref}

\newcommand{\re}{\text{\rm Re\,}}
\newcommand{\im}{\text{\rm Im\,}}

\newcommand{\br}{{\mathbb{R}}}  

\newcommand{\bz}{{\mathbb{Z}}}
\newcommand{\zn}{{\mathbb{Z}^\nu}}
\newcommand{\bc}{{\mathbb{C}}}

\renewcommand{\a}{\alpha}

\renewcommand{\l}{\lambda}

\newcommand{\s}{\sigma}
\renewcommand{\sp}{\sigma_p(J)}
\newcommand{\spp}{\sigma_p(H)}

\renewcommand{\th}{\theta}

\newcommand{\dd}{\Delta}

\renewcommand{\gg}{\Gamma}

\newcommand{\pp}{\Phi}
\newcommand{\ps}{\Psi}
\newcommand{\pspt}{{\Psi^+_{\tan\theta}}}
\newcommand{\psmt}{{\Psi^-_{\tan\theta}}}

\newcommand{\sh}{\#}

\newcommand{\lt}{\left}
\newcommand{\rt}{\right}

\newcommand{\dsp}{\displaystyle}
\newcommand{\kker}{{\textrm{Ker\,}}}

\allowdisplaybreaks
\numberwithin{equation}{section}

\newtheorem{theorem}{Theorem}[section]

\newtheorem{lemma}[theorem]{Lemma}

\theoremstyle{definition}

\begin{document}

\title[Lieb--Thirring inequalities]{Lieb--Thirring bounds\\ for complex Jacobi 
matrices} 
\author[]{L. Golinskii, S. Kupin}

\address{Mathematics Division, Institute for Low Temperature Physics and 
Engineering, 
47 Lenin ave., Kharkov 61103, Ukraine}
\email{leonid.golinskii@gmail.com}

\address{CMI, Universit\'e de Provence, 39, rue Joliot-Curie, 13453 
Marseille 
Cedex 13, France}
\email{kupin@cmi.univ-mrs.fr}

\date{\today}


\begin{abstract} 
We obtain various versions of classical Lieb--Thirring  bounds for 
one- and multi-dimensional complex Jacobi matrices. Our method is based 
on Fan--Mirski Lemma and seems to be fairly general.
\end{abstract}

\maketitle

\section*{Introduction}
\label{s0}
In a recent interesting paper \cite{la}, Frank--Laptev--Lieb--Seiringer obtain 
Lieb--Thirring bounds for a multidimensional Schr\"odinger operator 
$H=-\dd+V$ with a complex-valued potential. The authors say that they ``can also 
replace $-\dd$ in $H$ by any operator for 
which Lieb--Thirring bounds for real-valued potentials hold (but making 
the appropriate change in the exponent of $V$ on the right side of the 
inequalities)". The proposition seems to describe a complex-valued {\it diagonal 
perturbation}  of a  given self-adjoint operator. The method of the paper 
relies, though, on the special form of the unperturbed self-adjoint operator.

We move somewhat further in this direction. Namely, we prove 
Lieb--Thirring bounds for a {\it non-selfadjoint} operator $A$ provided 
the bounds for its real part $\re A=(A+A^*)/2$ are available. We neither assume 
$A$ to 
be a diagonal perturbation of a self-adjoint operator $A_0$, nor we use 
the specifics of $A_0$. 

The idea of the proof of the main result is very simple and transparent. First,  
Lieb--Thirring bounds for 
complex-valued Jacobi matrices are reduced to the self-adjoint case with 
the help of an elementary Fan--Mirski Lemma (see \cite[Proposition III.5.3]{bha}).
Then we use results of Hundertmark--Simon \cite{hs} for the self-adjoint 
Jacobi matrices. Since the latter paper contains ``small coupling'' and 
``large coupling'' bounds, we get pairs of estimates  for every case we 
consider.

More precisely, we are interested in the complex-valued symmetric Jacobi 
matrices of the form
\begin{equation}\label{e1}
J=J(\{a_k\},\{b_k\})=
\begin{bmatrix}
b_1&a_1&0&\ldots\\
a_1&b_2&a_2&\ldots \\
0&a_2&b_3&\ldots \\
\vdots&\vdots&\vdots&\ddots
\end{bmatrix}
\end{equation}
where $a_k, b_k\in\bc$. We assume $J$ to be a compact 
perturbation of  the free Jacobi matrix $J_0=J(\{1\},\{0\})$, or, equivalently, 
$\lim_{k\to+\infty} a_k=1$, $  
\lim_{k\to+\infty} b_k=0$. It is well-known, that in this situation
$\s_{ess}(J)=[-2,2]$. The point spectrum of $J$ is 
denoted by $\s_p(J)$; the eigenvalues $\l\in\s_p(J)$  have finite algebraic (and 
geometric) multiplicity, and the set of their limit points is on the interval [-2,2]  
(see, e.g., \cite[Lemma I.5.2]{gk}).
\begin{theorem}\label{t1} For $p\ge 1$,
\begin{eqnarray}
\qquad \sum_{\l\in\sp} (\re\l-2)^p_+&+&\sum_{\l\in\sp} (\re\l+2)^p_-\label{e3}\\
 &\le& c_p\lt(\sum_{k=1}^\infty |\re b_k|^{p+1/2}+4|\re a_k-1|^{p+1/2}\rt),
\nonumber\\
\qquad\sum_{\l\in\sp} (\re\l-2)^p_+&+&\sum_{\l\in\sp} (\re\l+2)^p_-\label{e2}\\
 &\le& 3^{p-1}\lt(\sum_{k=1}^\infty |\re b_k|^p+4|\re a_k-1|^p\rt),
\nonumber
\end{eqnarray}
where 
\begin{equation}\label{e4}
c_p=\frac 12 3^{p-1/2}\frac{\gg(p+1)}{\gg(p+3/2)}\frac{\gg(2)}{\gg(3/2)}.
\end{equation}
\end{theorem}
Above, $x_+=\max\{x,0\}$, $x_-=-\min\{x,0\}$ for $x\in \br$, so 
$$ x=x_+-x_-, \quad |x|=x_++x_-, \quad (-x)_+=x_-. $$ 

Note that Theorem \ref{t1} is a particular case of
more general results, Theorems \ref{t02} and \ref{t2}, of the same spirit.

The paper is organized in the following way. Section \ref{s1} contains the proof 
of the above theorem along with a number of other results for one-dimensional Jacobi 
matrices. 
Following the pattern of \cite{la}, we also get estimates on single eigenvalues 
for the complex Jacobi matrices. Similar theorems on multidimensional Jacobi 
matrices are in Section \ref{s2}.
Actually, the multidimensional results obviously give the estimates of Section 
\ref{s1}. Nevertheless, we prefer to state the bounds for the one-dimensional 
case explicitly.

We also mention that the proofs of the paper go through for general complex Jacobi 
matrices, not necessarily symmetric ones. Related problems concerning the geometry and
location of the discrete spectrum for such 
matrices are studied in \cite{eg1,eg2,eg3}.

\section{Lieb--Thirring bounds for one-dimensional Jacobi matrices}\label{s1}

\begin{figure}
\begin{center}
\includegraphics[height=8cm]{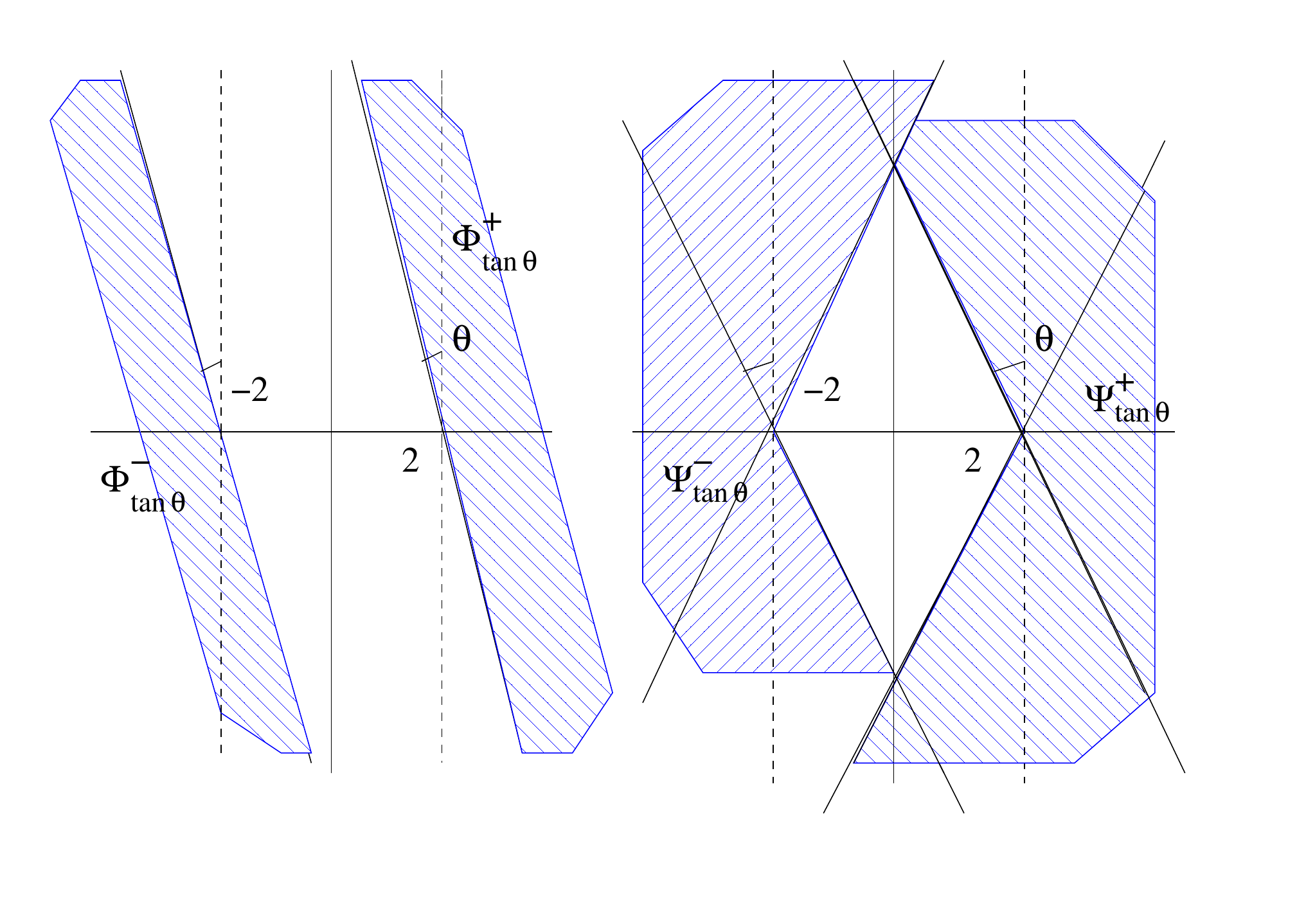}
\end{center}
\vspace{-1cm}
\caption{}\label{f1}
\end{figure}

Let $J=J(\{a_k\},\{b_k\})$ be a complex Jacobi matrix \eqref{e1}, 
and $\sp=\{\l_j\}$ its point spectrum which consists of eigenvalues of finite 
algebraic multiplicity. For $\a\in\br$, we introduce the functions
\begin{equation}\label{e101}
f^+_\a(\l)=(\re\l-2)+\a\,\im\l, \qquad f^-_\a(\l)=-(\re\l+2)-\a\,\im\l,
\end{equation}
and the half-planes
\begin{equation}\label{e102}
\pp^+_\a=\{\l: f^+_\a(\l)> 0\},\qquad  \pp^-_\a=\{\l: f^-_\a(\l)> 0\}.
\end{equation}
It is clear that $\l\in\pp_\a^- \Leftrightarrow -\l\in\pp_\a^+$.  Define also the angles
$$ \ps^+_\a=\pp^+_\a\cup\pp^+_{-\a}, \qquad \ps_\a^-=\pp_\a^-\cup\pp_{-\a}^-. $$
For $\a=\tan\th, -\pi/2<\th<\pi/2$, the regions are represented on Figure \ref{f1}.

We will be particularly concerned with the parts of the point spectrum $\sp$, lying in 
$\pp^\pm_\a$
and $\ps^\pm_\a$. We put   
$$
\s^\pm_p(J)=\{\l^\pm_{\a,j}\}=\sp\cap\pp^\pm_\a,
$$
and label the eigenvalues $\{\l^\pm_{\a,j}\}$ so that  
\begin{equation}\label{e103}
f^\pm_\a(\l^\pm_{\a,1})\ge f^\pm_\a(\l^\pm_{\a,2})\ge\ldots>0,\quad 
f^\pm_\a(\l^\pm_{\a,j})\searrow 0.
\end{equation}
The enumerating takes into account the multiplicities $l^\pm_j$ of $\l^\pm_{\a,j}$'s, so 
we have 
$$
\l^\pm_1=\ldots =\l^\pm_{l^\pm_1},\ \l^\pm_{l_1^\pm+1}=\ldots 
=\l^\pm_{l^\pm_1+l^\pm_2},\  \ldots
$$  
For instance,  we get for $\a=0$
\begin{eqnarray*}
f^\pm_0(\l)&=&\pm(\re\l\mp 2),\quad \pp^\pm_0=\{\l: \pm(\re\l\mp 2)>0\},\\
 \s^\pm_p(J)&=&\sp\cap\{\l: \pm(\re\l\mp 2)>0\}.
\end{eqnarray*}
Clearly, $\l\in \s_p^-(J) \Leftrightarrow -\l\in \s_p^+(-J)$.

Furthermore, in the notation 
$\re J=(J+J^*)/2$, $\im J=(J-J^*)/(2i)$, let 
$$ J_\a=\re J+\a\im J=J(\{\re a_k+\a\im a_k\},\{\re b_k+\a\im b_k\}) $$ 
be a {\it real self-adjoint} Jacobi matrix, and 
$$
\s_p(J_\a)=\s_p^-(J_\a)\cup\s_p^+(J_\a)=\{\mu^-_{\a,j}\}\cup\{\mu^+_{\a,j}\}
$$ 
be the set of its eigenvalues off the essential 
spectrum [-2,2], labelled as 
$$
\mu^-_{\a,1}<\mu^-_{\a,2}<\ldots<-2,\quad \mu^+_{\a,1}>\mu^+_{\a,2}>\ldots>2,
$$
and $\lim_{n\to\infty} \mu^\pm_{\a,n}=\pm2$.
In the case when one of the four numbers 
$$ l^\pm=\sum l^\pm_j=\sh\{\l^\pm_{\a,j}\}, \quad 
m^\pm=\sh\{\mu^\pm_j\} $$ is finite, the natural convention is that 
$\l^\pm_{\a,j}=\pm2$,\ $\mu^\pm_{\a,k}=\pm2$ for $j>l^\pm$ and $k>m^\pm$, respectively.
Observe that $\s^\pm_p(J)$, $l^\pm$, $m^\pm$ actually depend on $\a$, but we do not write 
this dependence to keep the notation reasonably simple.

\begin{lemma}\label{l1}  We have for $\a\in \br$ and $n=1,2,\ldots$
\begin{eqnarray}
\sum^n_{j=1} ((\re\l^+_{\a,j}-2)+\a \im\l^+_{\a,j})&\le&\sum^n_{j=1} 
(\mu^+_{\a,j}-2),\label{e301}\\
\sum^n_{j=1} ((\re\l^-_{\a,j}+2)+\a \im\l^-_{\a,j})&\ge&\sum^n_{j=1} (\mu^-_{\a,j}+2).  
\label{e302}
\end{eqnarray}
\end{lemma}

\begin{proof} The proof is a combination of the Fan--Mirski Lemma, the classical variational 
principle for eigenvalues, and 
elementary properties of a Schur basis for invariant subspaces of an operator \cite[Chapter 
1]{gk},
\cite{bha}. The argument is essentially finite dimensional, and is implicit in \cite[Chapter III]{bha}.

We will prove the first inequality. The second one, (\ref{e302}), is (\ref{e301}), applied 
to $-J$. 
We denote by $\{\nu^+_k\}$ the eigenvalues of $J$ in $\ps_\a$ {\it without} taking into 
account their multiplicities. More precisely, $\{\nu^+_k\}=
\{\l^+_{\a,j}\}$ as point sets, and
$$
f^+_\a(\nu^+_1)\ge f^+_\a(\nu^+_2)\ge\ldots>0,\quad f^+_\a(\nu^+_k)\searrow 0,
$$
but 
$$ \dsp \nu^+_1=\l^+_{\a,1}, \quad \nu^+_k=\l^+_{\a, \ \sum^{k-1}_{j=1}l^+_j +1} $$ 
for $k>1$. In 
particular, the corresponding root subspaces of $J$ are $H(\nu^+_k)=\kker 
(J-\nu^+_k I)^{l^+_k}$ and $\dim H(\nu^+_k)<+\infty$. 

For the transparency of the exposition, we assume that the {\it geometric 
multiplicity} of each $\nu^+_k$ {\it is one}. In this case $\dim H(\nu^+_k)=l^+_k$. 
The general situation is treated similarly. 

So, let $n$ be an 
arbitrary positive integer. We distinguish two cases: $n\le l^+$ and $n>l^+,\  
l^+<+\infty$. For the first one, there exists a unique $k_0=k_0(n)$ such that
$$
n=\sum^{k_0-1}_{k=1} n^+_k +n',\quad 0<n'\le l^+_{k_0}. 
$$
Put $H'(k_0)=\kker (J-\nu^+_{k_0} I)^{n'}$ and consider the direct sum
$$
\dsp H(n)=H(\nu_1^+)\stackrel{\cdot}{+}H(\nu_2^+)\stackrel{\cdot}{+}\ldots
\stackrel{\cdot}{+}H(\nu_{k_0-1}^+)
\stackrel{\cdot}{+}H'(k_0).
$$
Since the sum is direct and the summands are of finite dimension, we see that 
the subspace is closed, and  $\dim H(n)=n$. It is obvious that $H(n)$ is an 
invariant subspaces for $J$, and, by construction,
$\s_p(J(n))=\{\nu^+_k\}_{1\le k\le k_0},$
where $J(n)=J\big |_{H(n)}$.

We  now choose the Schur basis $\{x_j\}_{1\le j\le n}$ in $H(n)$. 
The system  $\{x_j\}$ by definition has the following 
properties:
\begin{itemize}
\item[--] it is orthonormal;
\item[--] for any $m\le n$, the linear span of $\{x_j\}_{1\le j\le m}$ is exactly 
$H(m)$;
\item[--] 
\quad $\dsp J x_j=J(n) x_j=\l^+_{\a,j} x_j+ \sum^{j-1}_{i=1} a_{j,i} x_i$
\end{itemize}
with some coefficients $\{a_{j,i}\}$. This means of course that the matrix of $J(n)$  
with respect to $\{x_j\}$ is upper-triangular.
 
Consequently, 
$$ \l^+_{\a,j}=(Jx_j,x_j), \qquad \bar \l^+_{\a,j}=(J^*x_j,x_j), $$ 
and we have $\re\l^+_{\a,j}=(\re J x_j,x_j)$, $\im\l^+_{\a,j}=(\im J x_j,x_j)$. So, 
$$
\re\l^+_{\a,j}+\a\,\im\l^+_{\a,j}=(J_\a x_j,x_j),
$$ 
and we continue as
\begin{equation}
\sum^n_{j=1} \re\l^+_{\a,j}+\a\,\im\l^+_{\a,j}=\sum^n_{j=1} (J_\a x_j,x_j) \label{e23}
\le\sup_{\{y_j\}_{1\le j\le n}\subset H} \sum^n_{j=1} (J_\a y_j,y_j),
\end{equation}
where the supremum is taken over all orthonormal systems $\{y_j\}_{1\le j\le n}$ 
in $H$. The operator $J_\a$ is self-adjoint and  its spectrum is 
described in the beginning of the current section. The spectral theorem says 
that
$$
J_\a=J_\a^\perp\oplus(J_\a)^+_p,\quad H=H^\perp\oplus H^+_p,
$$
where $H^\perp, H^+_p$ are reducing spectral subspaces associated to 
$\s^\perp(J_\a)=\s^-_p(J_\a)\cup\s_{ess}(J_\a)$ and $\s^+_p(J_\a)$, respectively. Obviously, 
$((J_\a)^+_px,x) \ge 2$, $(J^\perp_\a y,y)\le 2$, for $x\in H^+_p$, $y\in H^\perp$, and 
$||x||=||y||=1$. We proceed as
\begin{eqnarray*}
\textrm{LHS\ of\ } 
\eqref{e23}&\le&\dsp\max_{\{x_j\}_{1\le j\le\min\{n,m_+\}}\subset H^+_p}
\sum_j (J_\a x_j,x_j)\\
&+&\dsp\sup_{\{x_j\}_{\min\{n,m_+\}+1 \le j\le n}\subset H^\perp}
\sum_j (J_\a x_j,x_j)\\
&\le&\sum^{\min\{n,m_+\}}_{j=1}\mu^+_{\a, j}+2(n-m_+)_+,
\end{eqnarray*}
which is precisely \eqref{e301} under the notational convention made just before 
the lemma. 

When $n>l^+$, we have $\l^+_{\a,j}=2$ and $\mu^+_{\a,j}\ge2$ for 
$j>l^+$. So,
$$
\sum^n_{j=1}(\re\l^+_{\a,j}+\a\,\im\l^+_{\a,j})=
\sum^{l^+}_{j=1}(\re\l^+_{\a,j}+\a\,\im\l^+_{\a,j})+\sum^n_{j=l^++1}2
\le\sum^n_{j=1}\mu^+_{\a,j}.
$$
The proof is complete.
\end{proof}

\begin{lemma}\label{l2} For $p\ge 1$ and any $n=1,2,\ldots$,
\begin{equation}\label{e33}
\sum^n_{j=1}((\re \l^\pm_{\a,j}\mp 
2)+\a\,\im\l^\pm_{\a,j})^p_\pm\le\sum^n_{j=1}(\mu^\pm_{\a,j} \mp 2)^p_\pm. 
\end{equation}
Consequently,  
\begin{eqnarray}
\quad &&\sum^n_{j=1}((\re \l^+_{\a,j}-2)+\a\,\im\l^+_{\a,j})^p_++\sum^n_{j=1} 
((\re\l^-_{\a,j}+2)+\a\,\im\l^-_{\a,j})^p_-  \label{e34}\\
&\le&\sum^n_{k=1} |\mu^+_{\a,k}-2|^p+|\mu^-_{\a,k}+2|^p. \nonumber
\end{eqnarray}
\end{lemma}
\begin{proof}
Having \eqref{e301} at hand, the well-known Weyl's lemma \cite[Chapter 2]{gk}, \cite{ha} 
claims that for any non-decreasing convex function $g$,
$$
\sum^n_{j=1} g((\re\l^+_{\a,j}-2)+\a\,\im\l^+_{\a,j})
\le\sum^n_{j=1} g(\mu^+_{\a,j}-2),
$$
and the function $g(x)=x^p_+$, $p\ge1$, gives the first inequality in 
\eqref{e33}. We use \eqref{e302} instead of \eqref{e301} to prove the second one.  
Taking the sum of both bounds \eqref{e33}, we get \eqref{e34}.
\end{proof}

\begin{theorem}\label{t01} For $\a\in\br$ and $p\ge 1$,
\begin{eqnarray}
\label{e37} &&\sum_{j=1}^\infty ((\re \l^+_{\a,j}-2)+\a\,\im\l^+_{\a,j})^p_+ 
+\sum_{j=1}^\infty 
((\re\l^-_{\a,j}+2)+\a\,\im\l^-_{\a,j})^p_-  \\
&\le&c_p \lt(\sum_{k=1}^\infty |\re b_k+\a\,\im b_k|^{p+1/2}+
4|\re a_k-1+\a\,\im a_k|^{p+1/2}\rt), \nonumber\\
 \label{e38} && \sum_{j=1}^\infty ((\re \l^+_{\a,j}-2)+\a\,\im\l^+_{\a,j})^p_+ 
+\sum_{j=1}^\infty 
((\re\l^-_{\a,j}+2)+\a\,\im\l^-_{\a,j})^p_- \\
&\le&3^{p-1}\lt(\sum_{k=1}^\infty |\re b_k+\a\,\im b_k|^{p}+
4|\re a_k-1+\a\,\im a_k|^{p}\rt). \nonumber
\end{eqnarray}
\end{theorem}

\begin{proof} 
Fix $\a\in\br$ and assume that the RHS in \eqref{e37}, \eqref{e38} are finite. 
Then Theorems 2 
and 4 from \cite{hs} applied to the {\it self-adjoint} Jacobi matrix $J_\a$, 
give the desired bounds for $\sum_j |\mu^+_{\a,j}-2|^p+|\mu^-_{\a,j}+2|^p$.  The rest is 
\eqref{e34} with $n$ going to infinity.  
\end{proof}

It is clear that Theorem \ref{t1} is precisely \eqref{e37}, \eqref{e38} with $\a=0$.

The following result deals with the eigenvalues of $J$ in $\ps_\a^{\pm}$.

\begin{theorem}\label{t02} For $\a\ge 0$ and $p\ge 1$,
\begin{eqnarray}
\label{e39} && \sum_{\l\in\sp} ((\re \l -2)+\a\,|\im\l|)^p_+ 
+\sum_{\l\in\sp} 
((\re\l+2)-\a\,|\im\l|)^p_-  \\
&\le&c_p \Big(\sum_{k=1}^\infty |\re b_k+\a\,\im b_k|^{p+1/2}+
4|\re a_k-1+\a\,\im a_k|^{p+1/2} \nonumber\\
&+&\sum_{k=1}^\infty |\re b_k-\a\,\im b_k|^{p+1/2}+
4|\re a_k-1-\a\,\im a_k|^{p+1/2}\Big), \nonumber\\
 \label{e40} && \sum_{\l\in\sp} ((\re \l -2)+\a\,|\im\l|)^p_+ 
+\sum_{\l\in\sp} 
((\re\l+2)-\a\,|\im\l|)^p_-  \\
&\le&3^{p-1} \Big(\sum_{k=1}^\infty |\re b_k+\a\,\im b_k|^{p}+
4|\re a_k-1+\a\,\im a_k|^{p} \nonumber\\
&+&\sum_{k=1}^\infty |\re b_k-\a\,\im b_k|^{p}+4|\re a_k-1-\a\,\im a_k|^{p}\Big). \nonumber
\end{eqnarray}
\end{theorem}

\begin{proof}  
Let $\a\ge 0$. Bound \eqref{e37} implies that
\begin{eqnarray*}
&& \hspace{-1cm}\sum_{j:\ \im\l^+_{\a,j}\ge 0} ((\re\l^+_{\a,j}-2)+\a\,\im\l^+_{\a,j})^p_+ +
\sum_{j:\ \im\l^-_{\a,j}\le 0}
((\re\l^-_{\a,j}+2)+\a\,\im\l^-_{\a,j})^p_-  \\
&\le&c_p \lt(\sum_{k=1}^\infty |\re b_k+\a\,\im b_k|^{p+1/2}+
4|\re a_k-1+\a\,\im a_k|^{p+1/2}\rt),
\end{eqnarray*}
and
\begin{eqnarray*}
&& \hspace{-1cm} \sum_{j:\ \im\l^+_{-\a,j}<0} ((\re \l^+_{-\a,j}-2)-\a\,\im\l^+_{-\a,j})^p_+ 
+\sum_{j:\ \im\l^-_{-\a,j}>0} 
((\re\l^-_{-\a,j}+2)\\
&-&\a\,\im\l^-_{-\a,j})^p_-  
\le c_p \lt(\sum_{k=1}^\infty |\re b_k-\a\,\im b_k|^{p+1/2}+4|\re a_k-1-\a\,\im a_k|^{p+1/2}\rt).
\end{eqnarray*}
Since 
$$
\sp\cap\ps^\pm_\a=\{\l^\pm_{\a,j}:\ \pm\im\l^\pm_{\a,j}\ge0\} 
\cup \{\l^\pm_{-\a,j}:\ \pm\im\l^\pm_{-\a,j}<0\}),
$$
we obtain \eqref{e39} adding these two bounds. 

Note that transition from $\a$ to $-\a$ in the above formulae is equivalent to transition
from $J$ to $J^*$.
\end{proof}

We can refine \eqref{e37}  with a bit more precise inequalities
\begin{eqnarray}\label{e52}
\quad &&\sum_j ((\re\l^\pm_{\a,j}\mp2)+\a\,\im\l^\pm_{\a,j})^p_\pm
\le c_p\big(\sum_{k=1}^\infty (\re b_k+\a\,\im b_k)_\pm^{p+1/2}\\
&+&2|\re a_k-1+\a\,\im a_k |^{p+1/2}\big), \nonumber
\end{eqnarray}
the same applies to \eqref{e38}, Theorems \ref{t1}, \ref{t02} and their multidimensional 
counterparts,  
(see the proof of Theorem 1 in \cite{hs}).  

The ``angular'' Lieb--Thirring bounds are now an easy consequence of the previous theorem.
\begin{theorem}\label{t2} Let $p\ge1$ and  $0\le\th<\pi/2$.  Then
\begin{eqnarray}
\label{e7} \sum_{\l\in\sp\cap\pspt} |\l-2|^p&+&\sum_{\l\in\sp\cap\psmt} 
|\l+2|^p\\
&\le&c^1_{p,\th}
\lt(\sum_{k=1}^\infty |b_k|^{p+1/2}+4|a_k-1|^{p+1/2}\rt),
\nonumber\\
\label{e6} \sum_{\l\in\sp\cap\pspt} |\l-2|^p&+&\sum_{\l\in\sp\cap\psmt} 
|\l+2|^p\\
&\le&c^2_{p,\th} \lt(\sum_{k=1}^\infty |b_k|^p+ 4|a_k-1|^p\rt),
\nonumber
\end{eqnarray}
where 
$$
c^1_{p,\th}=2^{p/2+5/4}(1+2\tan\th)^{p+1/2}c_p,\quad 
c^2_{p,\th}=3^{p-1} 2^{p/2+1}(1+2\tan\th)^{p/2},
$$ 
and  $c_p$ is $(\ref{e4})$.
\end{theorem}

\begin{proof}  We will prove \eqref{e7}; the proof of \eqref{e6} is similar. 
 
Given $\th$, $0\le\th<\pi/2$, we pick $\th_1$, $\th<\th_1<\pi/2$, which solves the 
equation $\tan \th_1=1+2\tan\th$.  Write \eqref{e39} with $\a=\tan\th_1$:
\begin{eqnarray}
\label{e53} &&\\
 && \hspace{-1.5cm} \sum_{\l\in\sp\cap\pspt}((\re \l-2)+\tan\th_1\,|\im\l|)^p_+ 
 +\sum_{\l\in\sp\cap\psmt} 
((\re\l+2)-\tan\th_1\,|\im\l|)^p_-  \nonumber\\
&\le&c_p \Big(\sum_{k=1}^\infty |\re b_k+\tan\th_1\,\im b_k|^{p+1/2}+4|\re a_k-1+\tan\th_1\,\im 
a_k|^{p+1/2} \nonumber\\
&+&\sum_{k=1}^\infty |\re b_k-\tan\th_1\,\im b_k|^{p+1/2}+4|\re a_k-1-\tan\th_1\,\im 
a_k|^{p+1/2}\Big). \nonumber
\end{eqnarray}
Since $1\le \tan\th_1$ and $a+b\le\sqrt{2(a^2+b^2)}$ for $a,b\ge0$, we have
$$
\textrm{LHS\ of\ } 
\eqref{e53}
\le 2^{1+(p+1/2)/2}\tan^{p+1/2}\th_1\, c_p
\lt(\sum_{k=1}^\infty |b_k|^{p+1/2}+4|a_k-1|^{p+1/2}\rt),
$$
which is precisely the RHS of \eqref{e7}. 

Next, in the LHS of \eqref{e53} put $\l=x+iy$. 
If $x-2\ge 0$, then $(x-2)+|y|\tan\th_1\ge|\l-2|$. If $x-2<0$, we get
\begin{eqnarray*}
(x-2)+|y|\tan\th_1 &=&(x-2)+|y|(1+2\tan\th)\\
&=&((x-2)+|y|\tan\th)+|y|(1+\tan\th)\\
&\ge&|y|+|y|\tan\th=|y|+(2-x)\ge|\l-2|, 
\end{eqnarray*}
where we used $(x-2)+|y|\tan\th\ge 0$ for $\l\in\pspt$. 
The second term in the LHS of \eqref{e53} is handled similarly. The theorem is proved.
\end{proof}

For self-adjoint Jacobi matrices $J$ the bounds for individual eigenvalues $\l(J)$ drop out 
immediately from
\eqref{e52}
\begin{eqnarray}
(\l(J)\mp 2)^p_\pm&\le& c_p\lt(\sum_{k=1}^\infty (b_k)_\pm^{p+1/2}+2 
|a_k-1|^{p+1/2}\rt), \label{e74}\\
(\l(J)\mp 2)^p_\pm&\le& 3^{p-1}\lt(\sum_{k=1}^\infty (b_k)_\pm^{p}+2 
|a_k-1|^{p}\rt),\label{e73}
\end{eqnarray}
These estimates, modified appropriately, hold in the non-selfadjoint case as well.

\begin{theorem}\label{t3} Let $p\ge 1$, $J=J(\{a_k\},\{b_k\})$ be a complex 
Jacobi matrix, and $\l(J)$ its eigenvalue. Then
\begin{eqnarray}
\label{e8}&&\\
(\re \l(J)\mp2)^p_\pm&\le& c_p\lt(\sum_{k=1}^\infty (\re b_k)_\pm^{p+1/2}+2|\re 
a_k-1|^{p+1/2}\rt), \nonumber\\
\qquad(\re \l(J)\mp2)^p_\pm&\le& 3^{p-1}\lt(\sum_{k=1}^\infty(\re b_k)_\pm^p+2|\re 
a_k-1|^{p}\rt). \nonumber
\end{eqnarray}
When $\re\l(J)<-2$ or $\re\l(J)>2$, we have
\begin{eqnarray}
\label{e91} &&\\
|\l(J)\mp2|^p&\le&2^{p/2+1/4}\, c_p \big(\sum_{k=1}^\infty|b_k|^{p+1/2}
+2|a_k-1|^{p+1/2}\big), \nonumber\\
|\l(J)\mp2|^p&\le&2^{p/2}\, 3^{p-1} \big(\sum_{k=1}^\infty |b_k|^p
+2|a_k-1|^p\big). \nonumber
\end{eqnarray}
Finally, when $-2\le\re\l(J)\le2$, $\l(J)\notin [-2,2]$, we have
\begin{eqnarray}
\label{e10} &&\\
|\l(J)\mp 2|^p&\le& c_p\, (1+2\tan\th)^{p+1/2}\lt(\sum_{k=1}^\infty |b_k|^{p+1/2}+
2|a_k-1|^{p+1/2}\rt), \nonumber\\
|\l(J)\mp 2|^p&\le& 3^{p-1}\, (1+2\tan\th)^p\lt(\sum_{k=1}^\infty |b_k|^p+ 
2|a_k-1|^p\rt), \nonumber \end{eqnarray}
where $\th$ depends on a particular choice of $\l(J)$.

\end{theorem}

\begin{proof} The bounds in \eqref{e8} and \eqref{e91} are obvious in view of \eqref{e52}
(with $\a=0$) and \eqref{e7}--\eqref{e6} (with $\th=0$), where a single term in the LHS is 
taken 
instead of the whole sum. As far as \eqref{e10} goes, we pick $\th$ in such a way that
$\l(J)\in\ps^{\pm}_{\tan\th}$, and Theorem \ref{t2} does the rest.
\end{proof}

\section{Lieb--Thirring estimates for multidimensional Jacobi 
matrices}\label{s2}

We start recalling the definition of a multidimensional Jacobi matrix acting on 
$l^2(\bz^\nu)$. Traditionally, the set of unordered pairs $b=(ij),\ i,j\in 
\bz^\nu,\ |i-j|=1$, will be called the set of bonds $B(\zn)$. For 
$u=\{u(n)\}_{n\in\zn}\in l^2(\zn)$, we define
$H=H(\{a_b\}_{b\in B(\zn)}, \{b(n)\}_{n\in\zn})$ as
\begin{eqnarray}\label{e9}
(Hu)(n)&=&\sum_{|n-m|=1} a_{(nm)}u(m) + b(n)u(n),\\
(H_0u)(n)&=&\sum_{|m-n|=1} u(m). \nonumber
\end{eqnarray}
where $a_b,b(n)\in\bc$. We suppose of course that $H$ is a compact perturbation 
of $H_0$, or, $\lim_{|n|\to+\infty} a_{(nm)}=1$ and $\lim_{|n|\to+\infty} 
b(n)=0$. Then obviously $\s_{ess}(H)=[-2\nu,2\nu]$, and $\spp$ forms a sequence 
converging to the interval.

One can immediately write down counterparts of {\it all} results of 
Section~\ref{s1} for multidimensional Jacobi matrices. For the sake of brevity, we will
illustrate this taking Theorem \ref{t1} as an example.

The following bounds are obtained in \cite[Section 5]{hs} for 
self-adjoint operators $H$ (that is, for $a_b>0, b(n)\in\br$) and $p\ge1$:
\begin{eqnarray*}
\sum_{\l\in\spp}(\l-2\nu)_+^p&+&\sum_{\l\in\spp}(\l+2\nu)_-^p\\
 &\le&2^\nu (2\nu+1)^{p+\nu/2-1}L^{cl}_{p\nu}\lt(\sum_n |b(n)|^{p+\nu/2}+ 
2\sum_b |a_b-1|^{p+\nu/2}\rt),\\
\sum_{\l\in\spp}(\l-2\nu)_+^p&+&\sum_{\l\in\spp}(\l+2\nu)_-^p\\
 &\le&(2\nu+1)^{p-1}\lt(\sum_n |b(n)|^p+2\sum_b |a_b-1|^p\rt).
\end{eqnarray*}
where
$$
L^{cl}_{p\nu}=2^{-\nu}\pi^{-\nu/2}\frac{\gg(p+1)}{\gg(p+\nu/2+1)}.
$$
\begin{theorem}\label{t4}
Let $H$ be  a multidimensional complex Jacobi matrix described in \eqref{e9} 
and $p\ge 1$. Then
\begin{eqnarray*}
\sum_{\l\in\spp}(\re\l-2\nu)_+^p&+&\sum_{\l\in\spp}(\re\l+2\nu)_-^p\\
 &\le&2^\nu (2\nu+1)^{p+\nu/2-1}L^{cl}_{p\nu}\Big(\sum_n |\re b(n)|^{p+\nu/2}\\
 &+&2\sum_b |\re a_b-1|^{p+\nu/2}\Big),\\
\sum_{\l\in\spp}(\re \l-2\nu)_+^p&+&\sum_{\l\in\spp}(\re \l+2\nu)_-^p\\
 &\le&(2\nu+1)^{p-1}\lt(\sum_n |\re b(n)|^p+2\sum_b |\re a_b-1|^p\rt).
\end{eqnarray*}
\end{theorem}

\end{document}